\documentclass[12pt]{article}
\usepackage{graphicx}
\usepackage{amsmath,bm}
\usepackage{amsmath}
\usepackage{amssymb}
\usepackage{amsfonts}
\usepackage{multirow}
\usepackage{amsthm}
\usepackage{natbib}

\usepackage{bm, enumerate,xcolor}
\usepackage{graphicx,multicol}
\usepackage{float}
\usepackage{multirow}
\usepackage{epstopdf}
\usepackage{algorithm}
\usepackage{algorithmic}

\usepackage[colorlinks,linkcolor=blue,anchorcolor=black,citecolor=blue,urlcolor=black]{hyperref}

\usepackage{caption}
\usepackage{subcaption}
\usepackage{tikz}
\usepackage{booktabs}
\usepackage{multirow}
\usepackage{longtable}
\usepackage{array}
\usepackage{makecell}

\newcommand{\blind}{0}

\usepackage{amsthm, amsmath, amsfonts, amssymb,mathrsfs}
\theoremstyle{plain}

\newtheorem{theorem}{Theorem}
\newtheorem{lemma}{Lemma}

\theoremstyle{definition}

\newtheorem{definition}{Definition}
\newtheorem{remark}{Remark}

\usepackage{bm, enumerate, xcolor}

\newcommand{\T}{{ \mathrm{\scriptscriptstyle T} }}

\newcommand{\E}{\mathrm{E}}
\newcommand{\var}{\mathrm{var}}
\newcommand{\tr}{\mathrm{tr}}

\newcommand{\ATE}{\mathrm{ATE}}
\newcommand{\SATE}{\mathrm{SATE}}
\newcommand{\ave}{\mathrm{ave}}
\newcommand{\bx}{\bm{x}}
\newcommand{\bbeta}{\bm{\beta}}

\newcommand{\bpi}{\bm{\pi}}
\newcommand{\bY}{\bm{Y}}

\newcommand{\berror}{\bm{e}}
\newcommand{\X}{{X}}

\newcommand{\bone}{\mathbf{1}}
\newcommand{\cov}{\rm cov}
\newcommand{\alloc}{\rm alloc}
\newcommand{\ang}{\rm ang}

\newcommand{\sur}{\rm sur}

\newcommand{\bzero}{\mathbf{0}}

\begin{document}

\def\spacingset#1{\renewcommand{\baselinestretch}%
{#1}\small\normalsize} \spacingset{1}


\if0\blind
{
  \title{\bf Minimum free energy randomized design to improve covariate balance}
  \author{Haolin Chen and Jun Yu\thanks{
    All the authors have equally contributed to this work. Authors are listed alphabetically by their surname. The corresponding author is Jun Yu (yujunbeta@bit.edu.cn).}\hspace{.2cm}\\
    School of Mathematics and Statistics, Beijing Institute of Technology\\100081 Beijing, China}
  \maketitle
} \fi

\if1\blind
{
  \bigskip
  \bigskip
  \bigskip
  \begin{center}
    {\LARGE\bf Minimum free energy randomized design to improve covariate balance}
\end{center}
  \medskip
} \fi

\bigskip
\begin{abstract}
``Block what you can and randomize what you cannot'' is the core principle for treatment effect estimation in randomized controlled trials.  
Although a wealth of allocation strategies has been developed, an explicit trade-off between the covariate balance achieved by blocking and the robustness guaranteed by randomization is seldom quantified.  
Motivated by the second law of thermodynamics, this work posits a new criterion that lowers the covariate imbalance while maximizing the entropy that quantifies contrast and allocation diversity. 
The resulting optimal strategy, termed the minimum free energy randomized design, is then derived, thereby formally achieving such a trade-off. 
To facilitate practical implementation, we further develop a computationally efficient dynamic allocation algorithm with theoretical guarantees.  
Using a finite-sample variance decomposition, the proposed randomization strategy is shown to control covariate imbalance while preventing unobserved heterogeneity from dominating the mean squared error, thus retaining minimax efficiency under the prescribed design constraints.  
Extensive numerical simulations demonstrate that our method achieves superior statistical efficiency and greater robustness than existing approaches.
\end{abstract}

\noindent%
\spacingset{1.6}
\clearpage

\section{Introduction}
\label{sec:intro}

The precise estimation of treatment effects, often obscured by experimental fluctuations, is a fundamental problem in statistics. 
Significant research has addressed this issue, particularly in studies that include concomitant pre-treatment covariates. 
In this scenario, the true effect is primarily masked by two sources: random error from lurking nuisance factors and the covariate imbalance due to allocation.

The principle, ``block what you can and randomize what you cannot", introduced by \cite{box1978statistics}, is a cornerstone of experimental design for obtaining precise estimates of treatment effects.  
There is a long-standing understanding of the benefits brought by randomization and blocking \citep{fisher1935design,montgomery2017design}.
To be precise,
randomization in the allocation not only validates the independent error assumption for model-based inference but also averages out the effects of lurking nuisance factors, thereby naturally benefiting design-based inference.
On the other hand, when the sources of nuisance variability are known and controllable, blocking is a critical technique for mitigating variability transmitted by nuisance factors. In summary,  randomization and blocking are two fundamental techniques broadly adapted to enhance the statistical efficiency of comparing treatment and control groups.
Great success has been achieved under the guidance of this criterion in the realms of agriculture \citep{cox1958planning}, industry \citep{montgomery2017design}, clinical trials \citep{rosenberger2015Randomization,Ma2026car}, social science \citep{duflo2007using,athey2017econometrics} and technology \citep{Larsen2024Statistical}.

The core principle of mitigating confounders transmitted from covariate imbalance dates back to \cite{fisher1925statistical}, who introduced blocking via complete randomization (CR) within covariate-based strata. This idea was naturally generalized to continuous covariates with methodologies focused on balancing concomitant observations \citep{cox1958planning}.
One popular approach is rerandomization \citep[ReM,][]{morgan2012rerandomization}, a methodology formally introduced to address covariate imbalance before an experiment begins. This technique iteratively draws random allocations, rejecting those where the covariate imbalance exceeds a predefined threshold, and repeating the randomization process until a balanced allocation is achieved.

Building upon the initial framework of ReM, researchers have developed this field along two main axes. The first concerns the proper imbalance measure. Subsequent proposals include but are not limited to ReM in tiers of covariates \citep[$\rm ReM_T$,][]{morgan2015rerandomization}, rank‑based ReM \citep{Johansson2020Rerandomization}, Ridge‑ReM \citep{branson2018ridge}, cluster ReM \citep{Lu2023Cluster}, PCA‑ReM \citep{zhang2024pca}, and Bayesian ReM \citep[ReB,][]{Liu2025ReB}. These works extended the imbalance measure to account for high dimensionality, collinearity, and varying covariate importance.
The second axis addresses computational issues to achieve the covariate balance. For instance, \cite{kallus2018optimal,kallus2021optimality} employ mixed integer programming to identify allocations with minimum or smallest‑K imbalance. 
\cite{Krieger2019Nearly} propose a greedy pair-switching algorithm. 
Innovations in improving the ReM efficiency have also been broadly studied, such as pair‑switching \citep{zhu2023pair},  two-stage ReM \citep{yang2023rejective}, best‑choice ReM \citep{Wang20251best}, offering alternative strategies to obtain balanced allocations without excessive sampling.
Beyond these methodological developments, theoretical work has examined the asymptotic properties of ReM based inference.  
See \cite{li2018asymptotic,wang2022rerandomization} as examples.

Despite the impressive progress that has been made, the central challenge, as crystallized by \cite{li2018asymptotic}, remains: reducing covariate imbalance improves estimation precision, yet overly strict ReM constrains the randomization space and thereby undermines the basis for statistical inference. This inherent conflict between blocking for optimality and randomization for validity defines a core tension.
A growing consensus now acknowledges that any viable ReM strategy must navigate this trade-off, weighing covariate balance against randomization diversity \citep{johansson2021optimal, kallus2021optimality}. This recognition has spurred a range of methodological responses, from sample-size-dependent thresholds \citep{wang2022rerandomization} to selection among near-optimal allocations \citep{kallus2021optimality}. 
To the best of our knowledge, there is no clear metric that quantifies the trade‑off between covariate imbalance and loss of randomization, and consequently, no principled criterion to guide researchers in designing ReM schemes that properly reconcile the two.

\textbf{Our contributions.} The article makes several contributions. 
First, we examine CR from two complementary perspectives: uniformity, which parallels the blocking philosophy in experimental designs, and unpredictability, which stems from randomization itself. 
We introduce angular entropy to measure the directional uniformity of allocation vectors on the hypersphere, and allocation entropy to capture the diversity of distinct allocations. 
Within a minimax framework, we establish explicit connections between the least favourable mean squared error (MSE) and two forms of entropy: angular entropy (Theorem~\ref{thm:orth}) and allocation entropy (Theorem~\ref{thm:robustness}), respectively. Second, we formalize a trade-off between covariate balance and randomization by drawing an analogy with the second law of thermodynamics, where equilibrium minimizes the free energy, a balance between internal energy (measured by the covariate imbalance) and disorder (measured by the two aforementioned entropies). 
We propose a minimal free energy randomization design (MFER), which penalizes average covariate imbalance and maximizes the strategy’s entropy, thereby protecting against inflated variance caused by unobserved confounders. (Section~\ref{subsec:criterion}).
Third, we derive the explicit allocation probability that attains minimal free energy (Theorem~\ref{thm:opt-pi}), enabling practitioners to directly resolve the imbalance-randomization trade-off. 
To reduce computational cost, we further develop a dynamic allocation algorithm that achieves this probability with theoretical guarantees (Algorithm~\ref{alg:rem-spin} and Theorem~\ref{thm:spin}).

A finite-sample decomposition of the sample average treatment effect \citep[SATE,][]{imbens2015causal} under MFER separates the total variance into components driven by covariate imbalance and by residual variation (Theorem~\ref{lem:decomp}).  
The imbalance-driven component is directly controlled by MFER through its allocation probabilities. 
Theorems~\ref{lem:covariate} and~\ref{thm:though} show that balance induced by MFER does not unduly jeopardize inference in the sense that the residual component remains bounded even under adversarial configurations.  
As a complement, we establish exact variance reduction under the classical super‑population model (Theorem~\ref{thm:eff}).
Numerical justification on both simulation and real-world examples confirm the advantage of the proposed MFER design.

\section{Problem setups}
\subsection{Potential outcomes and design-based inference}

Consider an $n$-subject experiment indexed by $1,\ldots,n$.
Experimenters assign $W=(W_1,\ldots,W_n)^\T  \in\{0,1\}^n$ to corresponding units, where $W_i=1$ if the $i$th unit has been assigned treatment and $W_i=0$ for control.
Following the classical potential outcome framework \citep{rubin1974estimating}, let $Y_i(1)$ and $Y_i(0)$ be the treatment and control potential outcomes, respectively.
The average treatment effect (ATE) we are interested in is therefore defined by $\ATE:=n^{-1}\sum_{i=1}^n (Y_i(1)-Y_i(0))$.

In design-based inference framework, statisticians need to draw an inference on ATE based on observations $Y_i=W_iY_i(1)+(1-W_i)Y_i(0)$ for $i=1,\ldots,n$. 
The potential outcomes $Y_i(1)$ and $Y_i(0)$ are assumed to be fixed, and the randomness of the observation comes from the allocation $W_i$.
The fundamental difficulty of causal inference is that researchers can observe at most one potential outcome per unit, so half of the potential outcomes are missing.
As the $\ATE$ cannot be directly observed, researchers usually estimate it by
\begin{equation*}
    \SATE:=\frac{1}{n_1}\sum_{i=1}^n W_iY_i(1)-\frac{1}{n_0}\sum_{i=1}^n (1-W_i)Y_i(0),
\end{equation*}
where $n_1=\sum_{i=1}^n W_i$ and $n_0=n-n_1$.
Clearly, the SATE is the difference between the sample means of the treatment and control groups.
This estimator is unbiased when the design $W$ does not depend on the potential outcomes $Y_i(0)$ and $Y_i(1)$ \citep{Neyman1990On,li2018asymptotic}. 

\subsection{Covariate balancing}\label{subsec:cir}

Before conducting an experiment, experimenters usually collect a pretreatment covariate vector $\bx_i\in\mathbb{R}^p$ for the $i$th unit.
Potential imbalances in $\bx$ can act as a confounder.
To alleviate this, an allocation $W$ with a small value of $\sum_{i=1}^n (2W_i-1)\bx_i$ is usually recommended in practice.
Otherwise, the SATE may be influenced by both the treatment effect and covariate imbalance.

The effect of the covariate imbalance can be more clearly illustrated after rewriting the potential outcomes as in \cite{morgan2012rerandomization}. 
By assuming the treatment effect is additive, one has 
\begin{equation}\label{eq:additive1}
    Y_i(W_i)=\tau W_i+\beta_0+\bbeta^\T\bx_i+e_i,
\end{equation}
where $\tau$ is the ATE, $\beta_0+\bbeta^\T\bx_i$ is the projection of $Y_i(0)$ onto the space spanned by $(1,\bx)$, and $e_i$ is a residual yielded by ordinary least square.
It should be emphasized that \eqref{eq:additive1} does not imply a linear relationship between $\bx$ and $Y(0)$, $Y(1)$. 
Instead, it just represents a linear projection of $(Y(0),Y(1))$ onto $(1,\bx)$. The SATE can be re-expressed as
\begin{equation}\label{eq:additive2}
    \SATE=\tau+\bbeta^\T(\bar\bx_T-\bar\bx_C)+\bar{e}_T-\bar{e}_C,
\end{equation}
where $\bar\bx_T=n_1^{-1}\sum_{i=1}^n W_i\bx_i$, $\bar\bx_C=n_0^{-1}\sum_{i=1}^n (1-W_i)\bx_i$, $\bar e_T=n_1^{-1}\sum_{i=1}^n W_ie_i$ and $\bar e_C=n_0^{-1}\sum_{i=1}^n (1-W_i)e_i$. Here, the subscripts $T$ and $C$ denote the treatment and control conditions, respectively.
The assumption of an additive causal effect is not necessary. We opt to use it here for clear transparency. 

As highlighted in \citet[Chapter 4]{cox1958planning}, covariate balancing is a form of blocking: by equalizing the covariate distributions across treatment and control groups, it reduces the between-group variance and thereby mitigates the estimation variance.
In practice, this is achieved by minimizing an imbalance measure $\mathcal{D}(W)$, for instance, the Mahalanobis distance $\mathcal{D}(W)=(\bar{\bx}_T-\bar{\bx}_C)^\T\{{\cov}(\bar{\bx}_T-\bar{\bx}_C)\}^{-1}(\bar{\bx}_T-\bar{\bx}_C)$.

\section{Free energy motivated randomization criterion}
The difficulty of choosing a randomized allocation reflects a trade-off between two objectives:
(a) randomization, which insulates the study from all confounding, both observed and unobserved, and thereby provides a robust basis for causal inference \citep{angrist2009mostly};
(b) covariate balance, which reduces the chance variation from ${\cov}(\bar{\bx}_T-\bar{\bx}_C)$ and thus yields more precise estimates.
The experimenter must therefore strike a balance between randomization and efficiency gained by enforcing balance on key covariates.

\subsection{Lessons from completely randomization}\label{subsec:entropy}
To capture the randomness in an allocation, we first provide a mathematical definition of random design strategy.

\begin{definition}
    Let $W_{(1)},\ldots, W_{(m)}$ be the competing treatment allocation vectors that could possibly be used in an experiment.
    A random design strategy $\bpi$ is a probability measure supported on the competing set $\{W_{(1)},\ldots, W_{(m)}\}$ such that the corresponding $n$-dimensional treatment allocation vector will be adopted in the experiment that is chosen at random by sampling from $\bpi$.
\end{definition}

Consider an experiment with $n_1$ units assigned to treatment and $n_0$ to control. Here, $\pi_j$ is the probability that $W_{(j)}$  is adopted. In CR,  the random design strategy corresponds to a discrete uniform probability measure supported on all possible $0,1$ combinations with $\sum_{i=1}^n W_{(j)i}=n_1$ for all $j$.
As for another extreme case, if there is only one desired treatment allocation vector, say $W_{(1)}$, the random design strategy becomes a Dirac measure on $W_{(1)}$.
Although the ReM procedure does not give an explicit form of $\bpi$, the underlying $\bpi$ has already been constructed by the rejection sampling. 

Now, we are ready to show the advantages of CR from both the ``blocking'' and ``randomization'' perspectives.
The general idea comes from game theory. 
More concretely, the ATE inference can be conceptualized as a game: the experimenter, aiming to minimize the MSE of the SATE by choosing the treatment allocation $W$, while an adversarial Nature selects the potential outcomes to maximize the same MSE.


Let $\hat{\tau}(\bpi)$ be the SATE  with the treatment assign vector $W$ generated from $\bpi$.
Denote the average potential outcomes $r_0Y_i(1)+r_1Y_i(0)$ by $Y_i^{\ave}$ with $r_0=n_0/n, r_1=1-r_0$.
The difference between $\hat{\tau}(\bpi)$ and $\tau$ can be further recast into
\begin{equation}\label{eq:tau=sate}
    \hat{\tau}(\bpi)-\tau=\frac{1}{nr_1r_0}\left(W-r_1\bone_n\right)^\T \bY_c^{\ave},
\end{equation}
where $\bone_n$ denotes the $n$-dimensional vector of ones and $\bY_c^{\ave}=(Y_1^{\ave}-\bar{Y}^{\ave},\ldots,Y_n^{\ave}-\bar{Y}^{\ave})$ with $\bar{Y}^{\ave}=n^{-1}\sum_{i=1}^n {Y}_i^{\ave}$.
The bias and MSE of $\hat{\tau}(\bpi)$ have the following forms:
\begin{align}
    \E_{\bpi}(\hat{\tau}(\bpi)-\tau) &= \frac{1}{nr_0r_1}\{\E_{\bpi}(W - r_1\bone_n)\}^\T \bY_c^{\ave},\label{eq:mean-tau}\\
    \E_{\bpi}\{(\hat{\tau}(\bpi)-\tau)^2\} &= \frac{1}{(nr_1r_0)^2}(\bY_c^{\ave})^\T \E_{\bpi}\{(W - r_1 \bone_n) (W - r_1\bone_n)^\T\} \bY_c^{\ave}. \label{eq:var-tau}
\end{align}

Denote the covariate matrix by $X \in \mathbb{R}^{n \times p}$.
The unbiasedness of SATE can be directly observed from \eqref{eq:mean-tau}.
\begin{lemma}\label{lem:1}
     Assume that $\|\bY_c^{\ave}\|^2/n<\infty$ and $W$ is independent of potential outcomes conditionally on $X$. If  
          both $W$ and $\bone_n-W$ lie in the support of $\bpi$ with equal probability, then $\E_{\bpi}(\hat{\tau}(\bpi)-\tau)=0$.
\end{lemma}
Lemma~\ref{lem:1} shows that unbiasedness requires no more than a balanced design with equal marginal allocation probabilities for each unit. The condition is remarkably weak: it merely demands that the allocation mechanism be symmetric with respect to swapping treatment and control. Any design that treats the two groups symmetrically yields an unbiased estimator, though researchers typically select the allocation with minimal imbalance. 

The principal advantage of CR lies not in bias elimination but in its effect on the MSE.
From the blocking perspective, CR  behaves like an orthogonal design: the centred allocation vector \(Z(W)=W-r_1\bone_n\) has a covariance matrix proportional to the identity on the \((n-1)\)-dimensional subspace orthogonal to \(\bone_n\).
As a result, for any two contrast vectors that are orthogonal to \(\bone_n\) and to each other, their inner products with \(Z(W)\) are uncorrelated.
This isotropy implies that the MSE of the SATE in \eqref{eq:var-tau}, depends solely on the squared length of the centred potential outcome vector \(\bY_c^{\ave}\) and not on its orientation.
Thus, CR places all linear contrasts on an equal footing so that every contrast is estimated with the same precision, yielding a minimax guarantee in which the worst‑case risk depends only on the magnitude, not the direction, of \(\bY_c^{\ave}\).

To quantify this advantage, we introduce the concept of angular entropy. Define the standardized allocation vector
$
\theta(W)={Z(W)}/{\|Z(W)\|}={(n r_0 r_1)^{-1/2}}{(W-r_1\bone_n)}$,
which lies on the unit hypersphere \(\mathbb{S}^{n-2}\). 
A randomization strategy \(\bpi\) induces a discrete distribution on \(\mathbb{S}^{n-2}\). 
Because isotropy ensures fair comparison across contrasts, the orthogonality property described above is reflected in the uniformity \citep{wang2018On}. To measure such uniformity, we borrow the concept of entropy from information theory. For mathematical rigour, we smooth the discrete distribution by convolution with a von Mises–Fisher kernel to obtain a continuous density \(Q_{\bpi}(\theta)\). 
In this continuous regime, we slightly abuse notation by dropping the explicit dependence on the discrete allocation $W$ and letting $\theta \in \mathbb{S}^{n-2}$ directly denote the continuous random vector under $Q_{\bpi}$. The Kullback–Leibler divergence from the uniform distribution on \(\mathbb{S}^{n-2}\) then reduces, up to an additive constant, to the negative angular entropy
\[
H_{\ang}(\bpi):=-\int_{\mathbb{S}^{n-2}} Q_{\bpi}(\theta)\log\bigl( Q_{\bpi}(\theta)\bigr)\,d\theta.
\]
Consequently, the mean squared error in \eqref{eq:var-tau} can be recast and approximated under the continuous measure as
\begin{equation*}
  \E_{\bpi}\{(\hat{\tau}(\bpi)-\tau)^2\} = \frac{1}{nr_1r_0}(\bY_c^{\ave})^\T \E_{\bpi}\{\theta(W) \theta(W)^\T\} \bY_c^{\ave}  
  \approx \frac{1}{nr_1r_0}(\bY_c^{\ave})^\T \E_{Q_{\bpi}}(\theta \theta^\T) \bY_c^{\ave}.  
\end{equation*}
Under some mild smoothness assumptions \citep{klemela2000estimation}, the approximation error vanishes as the kernel bandwidth tends to zero. 
For theoretical purposes, we therefore adopt the continuous representation when discussing angular entropy, regarding the discrepancy as negligible.
 The following theorem associates the estimation performance with the randomness of the design strategy measured by $H_{\ang}(\bpi)$. 

\begin{theorem}\label{thm:orth}
      Assume that $\|\bY_c^{\ave}\|^2/n<\infty$ and $W$ is independent of potential outcomes conditionally on $\X$. The following results hold.
      \begin{enumerate}[(1)]
          \item The least favourable MSE against $Q_{\bpi}$ induced by $\bpi$ satisfies 
          \begin{equation}\label{eq:worst-mse1}
           \max_{\bY_c^{\ave}} \frac{(\bY_c^{\ave})^\T \E_{Q_{\bpi}}(\theta \theta^\T) \bY_c^{\ave}}{r_1r_0\|\bY_c^\ave\|^2}\le \frac{1}{n-1} + \sqrt{\frac{1}{2} \Big( H_{\ang,\max} - H_{\ang}(\bpi) \Big)},
          \end{equation}
          where $H_{\ang,\max}$ denotes the maximum angular entropy on $\mathbb{S}^{n-2}$. 
          \item The maximum angular entropy is attained uniquely by the uniform distribution on \(\mathbb{S}^{n-2}\). 
      \end{enumerate}
  \end{theorem}

The upper bound in \eqref{eq:worst-mse1} provides a  characterization of the relationship between geometrical randomness and minimax robustness.  
To be precise, as long as the angular entropy \(H_{\ang}(\bpi)\) is kept sufficiently high, the worst-case risk cannot explode. 
Interestingly, the CR induced discrete uniform distribution equivalents to the continuous uniform distribution on $\mathbb{S}^{n-2}$ in terms of second moments, and thus achieves the theoretical minimax MSE.
Together with the results in Theorem~\ref{thm:orth}, one can conclude the advantage of CR from the blocking perspective that CR maximizes the angular entropy \(H_{\ang}(\bpi)\) over all possible \(\bpi\).

We now turn to analysing the benefits of ``randomization''. 
The virtue of randomization lies not in averaging over possibilities, but in denying Nature a single one to exploit. 
A deterministic allocation, as noted by \cite{blackwell1979theory}, is inherently fragile as it allows Nature to anticipate and exploit it. Randomization offers a defence by rendering the allocation unpredictable. 
A natural measure is the entropy over the set of allocations, $H_{\alloc}(\bpi) := -\sum_{j=1}^m \pi_j \log (\pi_j)$.
CR maximizes \(H_{\alloc}(\bpi)\) among all strategies on a given collection of allocations, while a deterministic strategy, a Dirac measure, minimizes it. 
The following theorem connects estimation precision to the degree of randomness captured by \(H_{\alloc}(\bpi)\).

 \begin{theorem}\label{thm:robustness}
      Under the same assumptions in Theorem~\ref{thm:orth}, the following results hold.
      \begin{enumerate}[(1)]
          \item The least favourable MSE against a randomized design  $\bpi$ satisfies 
          \begin{equation}\label{eq:worst-mse}
            \max_{\bY_c^{\ave}}\frac{\E_{\bpi}\{(\hat{\tau}(\bpi)-\tau)^2\}}{\|\bY_c^{\ave}\|^2/n}\ge \left(\frac{1}{r_0}+\frac{1}{r_1}\right)\exp(-H_{\alloc}(\bpi)),
          \end{equation}
          with the equality holds when $\bpi$ is a Dirac measure.
          \item If $H_{\alloc}(\bpi)$ attends its maximum, the least favourable MSE (defined on the left-hand side of \eqref{eq:worst-mse}) achieves its minimum. 
      \end{enumerate}
  \end{theorem}

  The first result reveals that the deterministic strategy corresponding to the smallest entropy is usually riskier than the randomized strategy with a larger entropy in adversarial configurations.
  The statistical intuition has been acknowledged in ReM literature, such as \cite{morgan2012rerandomization}, \cite{li2018asymptotic}.
The second result implies the CR is minimax optimal, which is consistent with \cite{wu1981robustness}, \cite{li1983minimaxity}, and \cite{waite2021minimax}. 

\begin{figure}[htbp]
    \centering\spacingset{1.0}
    \includegraphics[width=1.0\textwidth]{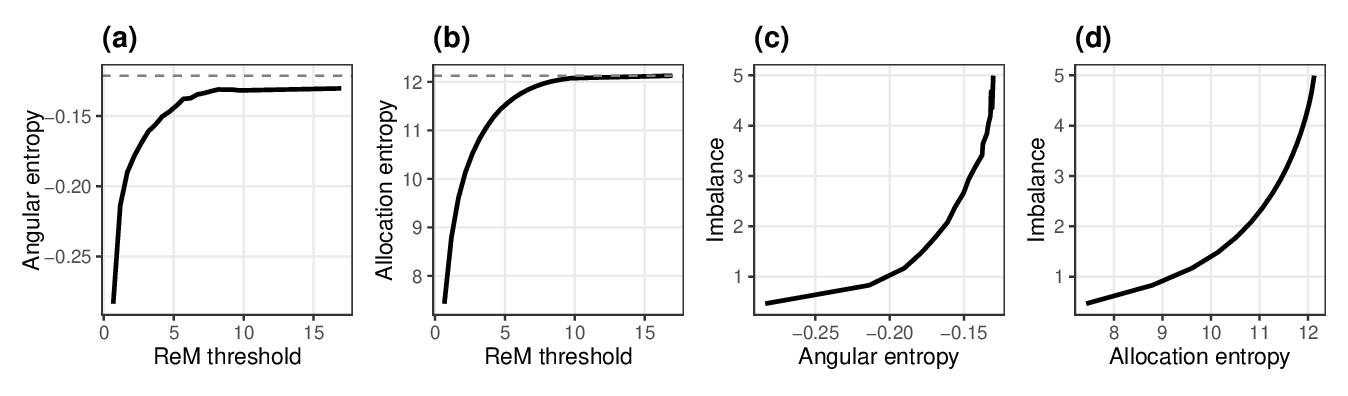}
    \caption{
    The relationships among the ReM threshold $a$, angular entropy $H_{\ang}$, allocation entropy $H_{\alloc}$, and expected covariate imbalance. The dashed lines in (a) and (b) represent the theoretical maximums achieved by CR.  
    }
    \label{fig:figure1}
\end{figure}

Interestingly, we note that the tolerance threshold \(a\) in ReM is intimately linked to both \(H_{\ang}(\bpi)\) and \(H_{\alloc}(\bpi)\). Figures \ref{fig:figure1}(a) and \ref{fig:figure1}(b) illustrate this connection in a simple setting with $n_1 = n_0 = 10$ and covariates generated from a multivariate normal distribution ${N}(\bzero_5, I_{5})$. One can easily figure out that as \(a\) tightens, both entropies decline. Figures \ref{fig:figure1}(c) and \ref{fig:figure1}(d) further reveal the trade‑off between these entropies and the expected covariate imbalance across different thresholds. Such relationships suggest that the two entropies may serve as a lens through which the choice of \(a\) can be understood. To be precise, a threshold should not be so stringent as to unduly sacrifice either form of entropy. This point implicitly speaks to the open problem raised by \cite{li2018asymptotic}.

A thermodynamic perspective elucidates the complementary roles of covariate balancing and randomization. As in a degenerate crystalline state, enforcing strict balance concentrates allocation vectors onto a low-dimensional subset of the allocation sphere. As a result, it sacrifices the independent variation in treatment contrast. Thereby, we will lose the ability to make fair comparisons in some directions. Angular entropy 
\(H_{\ang}(\bpi)\) mitigates this concentration by promoting directional uniformity of $W$, which directly controls the least-favourable MSE (Theorem~\ref{thm:orth}). 
 Concurrently, allocation entropy 
\(H_{\alloc}(\bpi)\) expands the support of the design, ensuring sufficient diversity of distinct treatment allocations and providing the lower bound on the least favourable MSE that underpins valid randomization inference (Theorem~\ref{thm:robustness}). 
Through the lens of thermodynamics, the crystalline concentration is stretched by angular entropy toward directional uniformity, while the restricted support is vaporized by allocation entropy into a diverse allocation ensemble. These complementary entropic effects collectively steer the allocation distribution toward the isotropic uniform measure on the sphere. Consequently, the contrast-invariance and minimax robustness characteristic of CR are recovered.

\subsection{Randomization criterion motivated by free energy}\label{subsec:criterion}

Building on the physical picture developed above, a desirable randomization strategy \(\bpi\) must navigate a fundamental tension that balances covariate (achieved through blocking) against the randomization captured by $H_{\ang}(\bpi)$ and $H_{\alloc}(\bpi)$. 
The second law of thermodynamics \citep{landau2013statistical} offers a compelling framework for understanding such trade-offs. 
Precisely, in a canonical ensemble held at constant temperature and volume, a system seeks to minimize its free energy \(F = U - TS\), where \(U\) represents internal energy and \(S\) represents entropy (the measure of disorder) \citep{jaynes1957information}. The second law dictates that at equilibrium, the system achieves an optimal compromise: it lowers its internal energy while simultaneously maximizing entropy, striking a balance such that \(F\) is minimized. This is Nature’s way of reconciling two competing tendencies through a single unifying principle.

Mapping this framework onto our problem, the \(n\) experimental units correspond to gas molecules. The expected covariate imbalance corresponds to internal energy \(U\), a design that aggressively balances covariates minimizes this ``energy'' but risks condensing into a rigid, low-entropy state. 
The two aforementioned entropies, \(H_{\alloc}(\bpi)\) and \(H_{\ang}(\bpi)\), together play the role of thermodynamic entropy \(S\): they measure the system’s disorder across both the number of distinct allocations and their contrast spread. 
Just as Nature balances \(U\) and \(S\) to minimize free energy, a robust randomization strategy must balance covariate balance against these two forms of entropy. 
From this analogy, our desired randomization design strategy $\bpi$ should be chosen so as to minimize the following ``free energy'' functional:
\begin{equation}\label{eq:fe}
F(\bpi) := \E_{\bpi}\{\mathcal{D}(W)\} - T_1 \cdot H_{\ang}(\bpi)-T_2\cdot H_{\alloc}(\bpi).
\end{equation}

The temperatures $T_1,\,T_2(\ge 0)$ here can be regarded as a 
belief in the variance reduction by projecting the potential outcomes on $X$.
When experimenters believe a perfect linear relationship between the potential outcomes and $X$, i.e., $Y(W)=\tau W+\beta_0+X\bbeta$, she can take $T_1,\,T_2$ to zero. In this scenario, minimizing the free energy becomes equivalent to minimizing the covariate imbalance alone. The resulting optimal design turns into the minimum imbalanced design studied in 
 \cite{kallus2018optimal}.
If experimenters believe the concomitant observations do not influence the inference, she can set either $T_{1}$ or $T_{2}$ to infinity, yielding the CR design.
Furthermore, through the lens of Lagrange multiplier, our criterion can also be explained as maximizing the entropy with the covariate imbalance being controlled at some certain level.

\section{Optimal randomization design and its implementation}
\label{sec:meth}

\subsection{Optimal randomization design strategy}

Lemma~\ref{lem:1} emphasis the value of using a balanced design such that $W$ and $\bone_n-W$ are exchangeable. 
An example in \cite{morgan2012rerandomization} also shows that SATE can be biased for ATE over ReM in an experiment with unequal $n_0$ and $n_1$.
Thus, we consider only the scenario $n_1=n_0=n/2$ for the rest of the paper.
 Let $m=\binom{n}{n/2}$ and $\{W_{(1)},\ldots,W_{(m)}\}$ be the set of treatment allocation vectors consisting of all possible $0,1$ combinations with $\sum_{i=1}^n W_{(j)i}=n/2$ for $j=1,\ldots,m$.

As $n$ increases, $m$ grows exponentially. 
This combinatorial explosion renders the exact evaluation of the kernel density estimator $Q_{\bpi}(\theta)$, and consequently the angular entropy $H_{\ang}(\bpi)$, computationally intractable. 
Thanks to the ensemble equivalence \citep{touchette2015ensemble}, the $Q_{\bpi}(\theta)$ constrained to the hypersphere $\mathbb{S}^{n-2}$(corresponding to the microcanonical ensemble) is suggested to be approximated by a Gaussian measure ${N}(\bzero_n, C)$ (corresponding to the canonical ensemble), where $C = \sum_j \pi_j \theta(W_{(j)}) \theta(W_{(j)})^\T$. 
Note that $\theta(W_{(j)})$ is orthogonal to $\bone_n$. 
We cannot directly approximate $H_{\ang}(\bpi)$ by $\det(C)$, which corresponds to the entropy of the Gaussian approximation (up to an additive constant), as $C$ is degenerate.
Instead, we opt to use $\sum_{l=1}^{n-1} \log \left(\lambda_l(C)\right)$, where $\lambda_l(C)$ is the $l$th largest eigenvalue of $C$. 
Substituting this approximation into \eqref{eq:fe} yields the surrogate free energy:
 \begin{equation}\label{eq:fe1}
 {F}_{\sur}(\bpi) := \E_{\bpi}\{\mathcal{D}(W)\} - T_1 \sum_{l=1}^{k} \log \left(\lambda_l(C)\right) - T_2 H_{\alloc}(\bpi),
 \end{equation}
 with $k=n-1$. 
 If a stringent covariate balance condition \(X^\T \theta(W) \approx \bzero\) is imposed, 
 the support approximately shrinks to the \((n-p-1)\)-dimensional orthogonal complement of $(\bone_n, X)$. 
 In this case, for numerical stability, we replace the summation bound in the second term from $k=n-1$ to \(k=n-p-1\).
 
Theorem~\ref{thm:opt-pi} characterizes the optimal random strategy induced by minimizing the surrogate free energy function~\eqref{eq:fe1}.

\begin{theorem}\label{thm:opt-pi}
The optimal solution to \(\min_{\bpi} F_{\sur}(\bpi)\) defined in \eqref{eq:fe1} exists and is unique.
The optimal $\bpi^*=(\pi^*_1,\ldots,\pi_m^*)^\T$ satisfies the self-consistent equation
\begin{equation}\label{eq:opt}
    \pi_j^* = \frac{1}{Z} 
    \exp\!\left( -\frac{\mathcal{D}(W_{(j)})}{T_2} 
          + \frac{T_1}{T_2}\,\theta(W_{(j)})^\T C^\dagger_*\,\theta(W_{(j)}) \right),
\end{equation}
where 
\( C_* = \sum_{j=1}^m \pi_j^*\,\theta(W_{(j)})\,\theta(W_{(j)})^\T \),
$
C^{\dagger}_* = \sum_{l=1}^{k} {\lambda_l}^{-1}\,{v}_l {v}_l^{\T}
$ with \(\lambda_1 \ge \cdots \ge \lambda_k > 0\) being the \(k\) largest eigenvalues of \(C_*\) and \(v_l\) being the corresponding orthonormal eigenvectors, and \( Z = \sum_{j=1}^m \exp( -{\mathcal{D}(W_{(j)})}/{T_2} 
          + {T_1}\,\theta(W_{(j)})^\T C^\dagger_*\,\theta(W_{(j)})/{T_2}) \)
is the normalization constant.
\end{theorem}

The entries of the optimal random design $\bpi^*$ in \eqref{eq:opt} can be factorized into the product of two terms $\pi_i^* \propto \exp\bigl(-{\mathcal{D}(W_{(i)})}/{T_2}\bigr) \cdot \exp( T_1\theta(W_{(i)})^\T C^\dagger_*\,\theta(W_{(i)})/T_2 )$. 
The first term is a covariate‑imbalance penalty that, exactly like ReM, excludes severely imbalanced allocations.  
The second term is a regularizer that quantifies the directional scarcity of \(\theta(W_{(j)})\) on \(\mathbb{S}^{n-2}\).  
When covariate balancing forces the allocation directions to collapse into a low‑dimensional subspace, the eigenvalues of the pseudoinverse \(C^\dagger_*\) diverge, generating a restoring force that pushes the design back towards isotropy.  
As the temperature ratio \(T_1/T_2\) dominates, the optimization forces all positive eigenvalues of the allocation covariance to be equal, recovering CR.  
Hence, by varying the temperatures \(T_1\) and \(T_2\), MFER generates a continuous family of designs that spans the spectrum from the observed efficiency obtained by covariate balance to the unobserved robustness of CR, moving towards a Pareto frontier between these two extremes.

\subsection{Practical implementation}\label{subsec:prac}

To render the computation tractable, we consider imbalance measures of the quadratic form
\begin{equation}\label{eq:imb-measure}
\mathcal{D}(W) = c\, (2W-1)^{\T} A (2W-1),
\end{equation}
where $c>0$ is a constant and $A$ is a positive semi-definite matrix.  
This class includes the squared Euclidean distance, the Mahalanobis distance, and the squared maximum mean discrepancy (MMD)  as special cases.   
They are obtained by taking $A$ to be $XX^{\T}$, $X(X^{\T}X)^{-1}X^{\T}$, and the kernel matrix $(K(\bx_i,\bx_j))_{i,j}$, respectively, with appropriate choices of $c$ (may depend on $n$).  
Here $K(\cdot,\cdot)$ is a user-specified reproducing kernel.  
Such measures have been widely adopted in the covariate balancing literature. See \cite{hu2012balancing}, \cite{morgan2012rerandomization}, \cite{kallus2018optimal,ma2024new} as examples.

Armed with the aforementioned imbalance measures, we now develop an efficient algorithm for MFER.  
Recall the optimal design \(\bpi^*\) is characterized by the self‑consistent mean‑field equation in Theorem~\ref{thm:opt-pi}. 
One can leverage a fixed‑point iteration to find the solution.
A direct optimization over the random design strategy \(\bpi\) is impossible for large \(n\) because the number of possible allocations \(m\) grows exponentially, making the parameter space prohibitively high‑dimensional.
The crucial observation that circumvents this difficulty is that the free energy functional depends on the $m$‑dimensional \(\bpi\) solely through the $n\times n$ matrix 
$C$.
Consequently, the effective number of degrees of freedom collapses from a combinatorial quantity, exponential in \(n\), to the dimension of \(C\), which remains small. Specifically, the optimization over the $m$-dimensional allocation simplex is reduced to a tractable problem over a compact convex set of $n \times n$ matrices. Exploiting the quadratic form of the imbalance measure, the first term of the free energy simplifies to \(\E_{\bpi}\{\mathcal{D}(W)\} = cn\,\tr(A C)\). Moreover, for a fixed \(C\), the allocation entropy can be maximized over all \(\bpi\) satisfying 
\(\E_{\bpi}\{\theta(W)\theta(W)^{\T}\}=C\), giving the entropy \(H(C)=\max_{\bpi: \E_{\bpi}\{\theta(W)\theta(W)^{\T}\}=C} H_{\alloc}(\bpi)\). Hence, the original free energy minimization is equivalent to 
\begin{equation}\label{eq:opt-c}
    \min_{C\,\in\,\mathcal{C}}\, \Bigl\{ cn\,\tr(A C) - T_1 \sum_{l=1}^{k} \log(\lambda_l(C)) - T_2 H(C) \Bigr\},
\end{equation}
where 
$\mathcal{C}$ denotes all convex combinations of $\theta(W)\theta(W)^\T$ with $W \in \{0,1\}^n, \bone_n^\T W = n/2$.
Since the objective function in \eqref{eq:opt-c} is strictly convex over the interior of the compact domain $\mathcal{C}$, the corresponding optimization problem can be efficiently optimized via natural gradient descent \citep{amari1998natural}.  

Based on the \(C_*\) obtained from \eqref{eq:opt-c}, the remaining task is to draw an allocation from the optimal design \(\bpi^*\).  
To circumvent the exponential cost of enumerating all $m$ allocations, we sample from $\bpi^*$ via a tailored Metropolis–Hastings scheme \citep{gelman2013BDA}.
To be precise, starting from a random allocation, the sampler repeatedly proposes a swap between two randomly chosen units, one from the treatment and one from the control.  
The swap is accepted with probability \(\min\!\big(1, \exp(-\Delta/T_2)\big)\), where \(\Delta\) is the change of \(\mathcal{D}(W) - T_1\,\theta(W)^\T C^\dagger_* \,\theta(W)\).  
The chain is run until equilibrium.
The complete procedure is summarized in Algorithm~\ref{alg:rem-spin}. 

\begin{algorithm}[!htbp]
\caption{Minimum free energy randomized (MFER) treatment assignment}
\label{alg:rem-spin}\spacingset{1.0}
\begin{algorithmic}[1]
\REQUIRE Covariate matrix $X$, imbalance measure $\mathcal{D}(W)$, temperatures $T_1,T_2$, stopping rule.
\STATE Obtain the matrix $C_*$ from \eqref{eq:opt-c}.
\STATE Initialize $W$ as a random permutation of $n/2$ ones and $n/2$ zeros.
\REPEAT
  \STATE Swap a randomly chosen treated unit and a control unit to form $W'$.
  \STATE $\Delta \gets \big\{\mathcal{D}(W') - T_1\,\theta(W')^\T C^\dagger_*\,\theta(W')-\mathcal{D}(W) + T_1\,\theta(W)^\T C^\dagger_*\,\theta(W)\big\}$; 
 \STATE Accept $W'$ with probability $\min\!\bigl(1,\;\exp\bigl(-\Delta/T_2\bigr)\bigr)$; if accepted, $W \gets W'$.
\UNTIL{stopping rule satisfied.}
\RETURN Final allocation $W^* \gets W$.
\end{algorithmic}
\end{algorithm}

Theorem~\ref{thm:spin} states that the final allocation $W^*$ can be regarded as a draw from $\bpi^*$.

\begin{theorem}\label{thm:spin}
    The sampling distribution of $W^*$ generated by Algorithm~\ref{alg:rem-spin} converges to the optimal randomize strategy $\bpi^*$ defined in \eqref{eq:opt} as the number of iterations tends to infinity.  
\end{theorem}

\section{Precision of the estimated treatment effect}
\label{subsec:verify}

Now we investigate the statistical behaviour of the SATE obtained by the randomized allocations from MFER under balanced designs. 

Let $H := X(X^\T X)^{-1}X^\T$ be the projection matrix onto the column space of $X$, and denote the corresponding residual vector by $\berror  := (e_1, \ldots, e_n)^\T= (I-H)\bY_c^{\ave}$.
The following theorem shows the variance decomposition of the $\SATE$.

\begin{theorem}\label{lem:decomp}
    Suppose $\bY(0),\bY(1)$ are independent of $W$ conditional on $X$,  $\berror\mid X {=} -\berror\mid X$ in distribution, and ${\cov}\big(H \bY_c^{\ave},\; (I-H)\bY_c^{\ave} | X\big) = 0$.
    Further assume the $T_1,T_2$ adopted in MFER  satisfy that $T_1/T_2<\infty$.
    Then, under Model \eqref{eq:additive1}, the SATE obtained from MFER satisfies that
    \begin{align}
       &\E_{\bpi^*,\bY_c^{\ave}}\{(\hat{\tau}(\bpi)-\tau)^2\}
       = \frac{4}{n}\E_{\bpi^*,\bY_c^{\ave}}\bigl\{(\bY_c^{\ave})^\T H  C_*  H \bY_c^{\ave}\bigr\}
       + \frac{4}{n}\E_{\bpi^*,\bY_c^{\ave}}\bigl\{ \berror^\T C_* \berror \bigr\},\label{eq:lem3-1}\\
       =&\frac{4}{n}\E_{\bpi^*,\bY_c^{\ave}}\bigl\{(\bY_c^{\ave})^\T H  C_*  H \bY_c^{\ave}\bigr\}+\frac{4}{n}\frac{\E_{\bY_c^{\ave}}\left[\E_{\bpi_{CR}}\left\{(\theta(W)^\T\berror)^2 \exp\!\left(B(W)  \right)\right\}\right]}{\E_{\bpi_{CR}}\left\{ \exp\!\left( B(W) \right)\right\}},\label{eq:lem3-2}
    \end{align}
    where $\bpi_{CR}$ is the random design strategy corresponds to CR, $\bpi^*$ and $C_*$ are given in Theorem~\ref{thm:opt-pi}, and $B(W)$ denotes $-{\mathcal{D}(W)}/{T_2} 
          + {T_1}\,\theta(W)^\T C^\dagger_*\,\theta(W)/T_2$.
\end{theorem}

For any strictly design-based inference, except CR,  the residual association between covariates and potential outcomes can always open a backdoor path through which Nature can inflate estimation error, for instance, by adjusting residual signs or within-stratum correlations. 
This is why ReM-based inference requires structural assumptions on the decomposition of potential outcomes. Specifically, \cite{morgan2012rerandomization} and \cite{li2018asymptotic} effectively assume that the potential outcomes can be split into a covariate-guided component and an orthogonal component that are independent or uncorrelated. 
Under this decomposition, randomization only shifts the former component while leaving the latter unaffected.
We adopt the same logic through our three assumptions, which formalize this separation for finite-sample inference. 
Crucially,  these assumptions are not essentially necessary. 
The covariance term between $H \bY_c^{\ave}$ and $(I-H)\bY_c^{\ave} $ admits an upper bound via standard inner-product inequalities. 
Hence, even without symmetry or uncorrelatedness, conservative inference remains feasible by replacing exact equalities with their upper bounds. 
Specifically, twice the right-hand side of \eqref{eq:lem3-2} yields valid error-rate control under no structural conditions on the residuals.

The result in \eqref{eq:lem3-1} separates covariate balance and residual variation, which enables us to take a close look at how covariate balance influences the MSE.
The first term corresponds to $\bbeta^\T{\cov}(\bar{\bx}_T-\bar{\bx}_C)\bbeta$ under Model~\eqref{eq:additive1} measuring the variance between groups. 
In the following, we show that this term is directly associated with the imbalance measure.
Without loss of generality, we take the imbalance measure $\mathcal{D}(W)$ in \eqref{eq:imb-measure}.
Let $R_0^2$ be the squared multiple correlation between $\bY_c^{\ave}$ and $X$; the 
 $S_{\bY_c^{\ave}}^2$ denote finite population variances of $\bY_c^{\ave}$. 
 
\begin{theorem}\label{lem:covariate}
   Assume the matrix \(A\) (used in the imbalance measure \(\mathcal{D}(W)\)) satisfies \(\operatorname{col}(X) \subseteq \operatorname{col}(A)\). For general $\bpi$, it holds
   \begin{equation}\label{eq:lem-c-11}
       \E_{\bpi,\bY_c^{\ave}}\bigl\{\bbeta^\T{\cov}(\bar{\bx}_T-\bar{\bx}_C)\bbeta\bigr\}\le \frac{4}{n}\frac{\sigma_{\max}(A^{\dagger}) (n-1)}{c n}\E_{\bY_c^{\ave}}\Bigl(R_0^2 S_{\bY_c^{\ave}}^2\Bigr)  \E_{\bpi}\Big\{\mathcal{D}(W)\Big\}, 
   \end{equation}
   where $\bbeta=(X^\T X)^{-1}X^\T \bY_c^{\ave}$, $\sigma_{\max}(A^{\dagger})$ is the maximum singular value of the generalized inverse of $A$.
   Moreover, if $\bpi=\bpi^*$ is given in Theorem~\ref{thm:opt-pi}, 
   \begin{equation}\label{eq:lem-c-1}
   \begin{split}
       \E_{\bpi^*,\bY_c^{\ave}}\bigl\{\bbeta^\T{\cov}(\bar{\bx}_T-\bar{\bx}_C)\bbeta\bigr\}=&\frac{4}{n}\E_{\bpi^*,\bY_c^{\ave}}\Bigl\{(\bY_c^{\ave})^\T H  C_*  H \bY_c^{\ave}\Bigr\}\\
       \le& \frac{4}{n}\frac{\sigma_{\max}(A^{\dagger}) (n-1)}{c n}\E_{\bY_c^{\ave}}\Bigl(R_0^2 S_{\bY_c^{\ave}}^2\Bigr)  \E_{\bpi^*}\Big\{\mathcal{D}(W)\Big\}.
   \end{split} 
   \end{equation}
\end{theorem}

Theorem~\ref{lem:covariate} shows that the component driven by covariate imbalance is bounded by a quantity proportional to the expected imbalance measure, providing a direct link between design parameters and estimation precision. 
The condition $\operatorname{col}(X) \subseteq \operatorname{col}(A)$ enables us to deal with a broad class of imbalanced measures as established in Section~\ref{subsec:prac}. 
For the Mahalanobis distance-based imbalance measure, 
$c = (n-1)/n$ and $\sigma_{\max}(A^{\dagger}) = 1$
which further simplified the bound on the inequality.

The second term in \eqref{eq:lem3-1} encapsulates the variance of the residual contrast, $\var(\bar{e}_T - \bar{e}_C)$. To investigate its potential variance inflation in a design-based analysis, we consider a structured adversarial setting. For any general random design $\bpi$, the variance of the residual contrast is strictly governed by its allocation covariance matrix, expressed as
$\var_{\bpi}(\bar{e}_T - \bar{e}_C) = {4}{n}^{-1} \berror^\T \E_{\bpi}\{\theta(W)\theta(W)^\T\} \berror.$
Let $\E_{\bpi}\{\theta(W)\theta(W)^\T\} = \sum_{l=1}^{n-1} \lambda_l v_l v_l^\T$ denote its spectral decomposition, with eigenvalues arranged in descending order $\lambda_1 \ge \cdots \ge \lambda_{n-1} \ge 0$. Under the covariate balance framework, it is naturally anticipated that the leading $q$ eigenvectors reside in the orthogonal complement of the column space of $(\bone_n, \X)$. In contrast to the minimax game detailed in Section~\ref{subsec:entropy}, we consider a more general adversarial model where Nature restricts the residual vector to this $q$-dimensional vulnerable subspace. Specifically, the residual is constructed as $\berror = \sum_{l=1}^q \omega_l v_l$, with the weight vector $(\omega_1, \ldots, \omega_q)$ uniformly distributed on the unit hypersphere $\mathbb{S}^{q-1}$.
The worst case discussed in Section~\ref{subsec:entropy} is the special case with $q=1$.
When the proposed MFER framework is adopted, the covariance $\E_{\bpi}\{\theta(W)\theta(W)^\T\}$ reduces to $C_*$, and the variance simplifies to ${4}{n}^{-1} \E_{\bY_c^{\ave}}(\berror^\T C_* \berror)$.
The following theorem evaluates the expected estimation risk $\var(\bar{e}_T-\bar{e}_C)$ under this adversarial model.
\begin{theorem}\label{thm:though}
    Suppose $\bY(0),\bY(1)$ are independent of $W$ conditional on $X$, and the first $n-q-1$ eigenvectors of $\E_{\bpi}\{\theta(W)\theta(W)^\T\}$ lie in the orthogonal complement of the column space of $(\bone_n,X)$. 
    If $\berror=\sum_{l=1}^q \omega_l v_{l}$,  and $(\omega_1,\ldots,\omega_q)$ is uniformly distributed on $\mathbb{S}^{q-1}$.
    Then, for any $q\le n-p-1$, it holds
    \begin{equation}
        \var(\bar{e}_T-\bar{e}_C)\le \frac{4}{n}\left(\frac{q+1}{q}\right)^{q+1}\frac{1}{\prod_{j=1}^{k}\lambda_i}\left(\frac{\tr({\E_{\bpi}\{\theta(W)\theta(W)^\T\}}(I-H))}{(k+1)}\right)^{k+1},
    \end{equation}
    where $k=n-p-1$.
\end{theorem}

The second condition of Theorem~\ref{thm:though} is used to guarantee that the residuals remain compatible with a design‑based inference framework. It holds under a general random design strategy $\bpi$.
When MFER  is applied with the Mahalanobis distance as the imbalance measure,
a trade‑off emerges directly from the identity \(\tr\{C_*(I_n-H)\}=1-\tr(HC_*)\), which governs the leading term in the numerator. Simple calculation yields that \(\E_{\bpi^*}\{\mathcal{D}(W)\}=(n-1)\tr(HC_*)\) (under Mahalanobis distance), enforcing covariate balance reduces \(\E_{\bpi^*}\{\mathcal{D}(W)\}\) and consequently increases the numerator through \(1-(n-1)^{-1}\E_{\bpi^*}\{\mathcal{D}(W)\}\). 
By rearranging the temperature factors, $F_{\sur}(\bpi)$ can also be interpreted as $\max  \sum_{l=1}^{k} \log \left(\lambda_l(C_*)\right)$ subject to prescribed levels of $\E_{\bpi}\{\mathcal{D}(W)\}$ and $H_{\alloc}(\bpi)$.
The proposed MFER appears in the denominator. 
One can expect our proposed MFER to lower the unfavourable MSE among all $\bpi$ with the same expected imbalance measure and allocation entropy.

Theorem~\ref{lem:decomp} not only provides a design‑based framework for disentangling covariate balance and residual variation, but also yields a super‑population insight. If $\bar{e}_T-\bar{e}_C$ is independent of the covariate contrasts $\bar{\bx}_T-\bar{\bx}_C$ under CR, hence of $C_*$ and $\mathcal{D}(W)$ as well, then the residual term in~\eqref{eq:lem3-2} simplifies to
\begin{align}
    &\frac{\E_{\bY_c^{\ave}}\left[\E_{\bpi_{CR}}\left\{(\theta(W)^\T\berror)^2 \exp\!\left( B(W) \right)\right\}\right]}{\E_{\bpi_{CR}}\left\{\exp\!\left( B(W) \right)\right\}}
    =\frac{\E_{\bpi_{CR},\bY_c^{\ave}}\left\{(\theta(W)^\T\berror)^2\right\}\E_{\bpi_{CR}}\left\{\exp\!\left( B(W) \right)\right\}}{\E_{\bpi_{CR}}\left\{\exp\!\left( B(W) \right)\right\}}\notag \\
          &=\E_{\bpi_{CR},\bY_c^{\ave}}\left\{(\theta(W)^\T\berror)^2\right\}=\E_{\bY_c^{\ave}}\left\{ (1-R_0^2)S_{\bY_c^{\ave}}^2\right\}.
\end{align}
 The residual contribution is therefore governed entirely by the predictive power of the covariates, and the allocation under MFER leaves the orthogonal residuals unaffected.

Now we are ready to show the variance reduction under the classical setup in \cite{morgan2012rerandomization}.

\begin{theorem}\label{thm:eff}
If (i) $X$ and $\bY^{\ave}$ are jointly normally distributed, and (ii) the treatment effect is additive. Let the imbalance measure be the  Mahalanobis distance. 
Suppose that there exists a parameter pair $(T_1^*, T_2^*)$ such that the corresponding MFER design satisfies the following bound:
\begin{equation}\label{eq:trace_bound}
    \E_{\bpi^*}\{\mathcal{D}(W)\} \le {v_a},
\end{equation}
for some constant $v_a >0$.
Then for any MFER strategy $\bpi^*$ given in Theorem~\ref{thm:opt-pi} with $T_2 \le T_2^*$ and $T_1/T_2=T_1^*/T_2^*$, the variance reduction ratio satisfies:
\begin{equation}
    {\rm VR}(\bpi^*) := \frac{\E_{\bpi^*,\bY_c^{\ave}}\{(\hat{\tau}(\bpi) - \tau)^2\}}{\E_{\bpi_{CR},\bY_c^{\ave}}\{(\hat{\tau}(\bpi) - \tau)^2\}} \le 1 - (1 - v_a)R_0^2.
\end{equation}
\end{theorem}

\begin{remark} \label{rmk:existence_of_parameters}
From the proof of Theorem~\ref{thm:eff}, $\E_{\bpi^*}\{\mathcal{D}(W)\}$ is a non-increasing function with respect to $T_2$ when the ratio $T_1/T_2$ is fixed or $T_1=0$.
The existence of $T_1^*,\,T_2^*$ can be guaranteed if the smallest imbalance measure among all possible candidates is less than $v_a$. 
In this case, one needs only specify $T_1^*=T_2^*=0$. (We define $0/0=0$ for simplicity).
 Based on the results of  \cite{Krieger2019Nearly} and \cite{ma2024new}, $\min_W \mathcal{D}(W)$ will converge to zero as $n \to \infty$ under very mild conditions. 
As $v_a > 0$ is a constant, the condition \eqref{eq:trace_bound} holds for sufficiently large $n$.
\end{remark}

{Analyzing the estimator in the classical ReM configuration of \cite{morgan2012rerandomization} is valuable, as it yields an exact, finite‑sample inference framework. In large samples, covariate imbalance decreases naturally even under CR.
While in small experiments, deliberate balancing becomes critical.
Theorem~\ref{thm:eff} shows that the proposed method is not less efficient than ReM, thus retaining the precision gains due to stringent covariate balance even when $n$ is limited.
The result relies on the independence between the randomization $W$ and the population errors $\berror$. While this independence holds by design under the super-population model, in finite samples, a minor residual dependence may exist. 
However, the reasoning may still plausibly extend to more general settings as approximate independence emerges asymptotically \citep{li2018asymptotic}. 
This super‑population perspective does not contradict the design‑based framework described in Theorem~\ref{thm:though}. 
Under a proper super‑population model, the randomness of the sampled potential outcomes $\bY^{\ave}$ helps average out residual variation, ensuring that the worst‑case configurations that drive the design‑based bounds are not inevitable in finite samples.}

{
\section{Numerical studies}
\label{sec:simulation}

In this section, we evaluate the finite sample performance of the proposed MFER method. Consider the potential outcome generated by Model \eqref{eq:additive1} with $\bbeta=2\cdot\bone_p$ and $\tau=2$. 
 The $\bx_1,\ldots, \bx_n$ are i.i.d. random vectors, where each component is independently drawn from a Student's t-distribution with degrees of freedom two ($t_2$).
 The results for other distributions (such as normal and log-normal) are quite similar, and thus we omit them for brevity.

To evaluate robust inference for estimating treatment effects, we consider the following three processes for generating the residual component.
To ensure algebraic orthogonality, we project the random vectors obtained from the following data generation process onto the orthogonal complement of the space spanned by $(\bone_n, X)$. The $\sigma$ presented in the following is changed case-by-case to keep $R_0^2$ fixed at 0.4.

\begin{description}
    \item[\texttt{R1}] (Independent error) The $e_1,\ldots, e_n$ are i.i.d. from normal distribution $N(0,\sigma^2)$.
    \item[\texttt{R2}] (Least favourable error) The least favourable case where $(e_1,\ldots, e_n)^\T=\sigma {v}_{\max}$ with $v_{\max}$ being the eigenvector corresponds to the maximum eigenvalue of $\E_{\bpi}\{(2W-1)(2W-1)^\T\}$.
    \item[\texttt{R3}] (Incomplete information game) Let $v_1,\ldots,v_q$ be the eigenvectors corresponds to the top-$5$ eigenvalue of $\E_{\bpi}\{(2W-1)(2W-1)^\T\}$. The residual $(e_1,\ldots, e_n)^\T=\sigma(\sum_{j=1}^5 \omega_j v_{j})$,  and $(\omega_1,\ldots,\omega_5)$ is uniformly distributed on $\mathbb{S}^{4}$. The scenario corresponds to the thought experiment we described in Theorem~\ref{thm:though}.
\end{description}

For each data generating process, we consider the following five randomization schemes: the proposed MFER (Algorithm~\ref{alg:rem-spin}), ReM,  Gram-Schmidt Walk design \citep[GSW,][]{harshaw2024balancing}, Matched Pairs \citep[MP,][]{greevy2004optimal}, and CR. We adopt the Mahalanobis distance as the primary covariate imbalance measure.
We highlight that \texttt{R2} and \texttt{R3} are not applicable under CR as its design matrix is a scalar multiple of the identity on the subspace orthogonal to $\bone_n$. Thus, the corresponding results are marked by `-'.

 All balance-adaptive designs, namely MFER, ReM, and GSW, are tuned  so that their expected imbalance $\E_{\bpi}\{\mathcal{D}(W)\}$ is approximately $30\%$ of that under CR, thus aligning covariate balance across these methods.
CR and MP offer no direct control over imbalance and are reported for reference.
Figure~\ref{fig:pareto_frontier} explores a wider range of imbalance levels, from $20\%$ to $100\%$ of the CR value, and confirms that the results are not sensitive to the particular choice of the $30\%$ threshold.
The sensitivity analyses in Section~\textcolor{blue}{S.2} further verify that the findings are robust to the choice of the algorithm  tuning parameters $T_1$ and $T_2$.

We initially evaluate a small-sample scenario ($n=20$) in which the exact allocation space can be enumerated, enabling precise calculation and verification of the expected imbalance levels. Subsequently, we scale the simulations to $n=100$ while fixing the $p/n$ ratio at 0.25 to assess the scalability and robustness of the proposed methods. All numerical studies are carried out with \texttt{R} (version 4.5.1) on an Apple Silicon M4 platform with 16 GB RAM.

Table~\ref{tab:results} presents the empirical (relative) MSE of SATE under different data generating processes (\texttt{R1}--\texttt{R3}). To ensure comparability, we use MFER as the benchmark and report each method's MSE relative to it. We make the following three observations. Firstly, covariate balancing improves the estimation efficiency compared with CR as the $R_0^2$ in our setups is away from zero, and thus reduces the randomization variance caused by the covariates' mean difference.
Secondly, for the independent residual (\texttt{R1}), all methods perform similarly (the difference between the three competitors is not greater than $5\%$) when the covariate imbalance is controlled at a comparable level. The phenomenon is characterized by the fact that different random design strategies do not affect the independence and distribution of random error. Thus, the only object for the experimenters to attend to is the imbalance measure.
Thirdly, the extreme balancing of the covariate may lead to an undesirable consequence, even in the least favourable scenario (\texttt{R2}), where experimenters play first and Nature plays second.
\texttt{R3} illustrates the case where some important confounders are not taken into account before random allocation. In this scenario, MFER is the least harmful strategy when the imbalance measure across strategies is controlled at a given value.

\begin{table}[!h]
\centering\spacingset{1.0}
\renewcommand{\arraystretch}{0.9}
\setlength{\tabcolsep}{12pt}
\caption{\label{tab:results}
Average imbalance (imb), max eigenvalue of $\E_{\bpi}\{(2W-1)(2W-1)^\T\}$ ($\lambda_{\max}$) and relative MSE under \texttt{R1}--\texttt{R3} among different methods. 
For CR, \texttt{R2} and \texttt{R3} are omitted (marked `-') since no least favourable direction exists. 
The CR and MP cannot control the imbalance and are reported for reference. 
The optimal values among the remaining three covariate-adaptive designs are highlighted in bold. }
\centering
\begin{tabular}[t]{llrrrrr}
\toprule
\multicolumn{4}{c}{ } & \multicolumn{3}{c}{MSE} \\
\cmidrule(l{3pt}r{3pt}){5-7}
\raisebox{3.25ex}[0pt][0pt]{ } & \raisebox{3.25ex}[0pt][0pt]{Method} & \raisebox{3.25ex}[0pt][0pt]{imb} & \raisebox{3.25ex}[0pt][0pt]{$\lambda_{\max}$} & \multicolumn{1}{c}{\raisebox{1.5ex}{\small \texttt{R1}}} & \multicolumn{1}{c}{\raisebox{1.5ex}{\small \texttt{R2}}} & \multicolumn{1}{c}{\raisebox{1.5ex}{\small \texttt{R3}}}\\[-2ex]
\midrule
 & MFER & 1.49 & \textbf{1.00} & 1.00 & \textbf{1.00} & \textbf{1.00}\\

 & ReM & 1.49 & 1.44 & \textbf{0.95} & 1.44 & 1.07\\

 & GSW & 1.49 & 1.68 & 0.98 & 1.68 & 1.14\\

 & MP & 2.39 & 1.14 & 1.09 & 1.17 & 1.28\\

\multirow{-5}{*}{\raggedright\arraybackslash $n = 20$} & CR & 5.00 & 0.60 & 1.05 & \multicolumn{1}{c}{-} & \multicolumn{1}{c}{-}\\
\midrule

 & MFER & 7.50 & \textbf{1.00} & \textbf{1.00} & \textbf{1.00} & \textbf{1.00}\\

 & ReM & 7.64 & 1.34 & 1.05 & 1.29 & 1.22\\

 & GSW & 7.49 & 1.23 & 1.01 & 1.22 & 1.16\\

 & MP & 15.66 & 1.25 & 1.21 & 1.26 & 1.43\\

\multirow{-5}{*}{\raggedright\arraybackslash $n = 100$} & CR & 25.02 & 0.67 & 1.20 & \multicolumn{1}{c}{-} & \multicolumn{1}{c}{-}\\
\bottomrule
\end{tabular}
\end{table}

Understanding the statistical behaviour of the proposed MFER under different levels of covariate imbalance is valuable. Figure~\ref{fig:pareto_frontier} reports the top two eigenvalues of $\E_{\bpi}\{(2W-1)(2W-1)^\T\}$  against different imbalance $\E_{\bpi}\{\mathcal{D}(W)\}$ under $n=20$, which reflect the potential risk caused by covariate balancing.
The results of CR are reported as a benchmark.
We choose $n=20$ for the following two reasons. Firstly, when the difference in covariate means naturally decreases as the sample size increases, the covariate balancing may not be the most important thing for causal inference \citep{morgan2012rerandomization}.
Secondly, the imbalance is easier to control when it is nearly the same, which makes the comparison fairer. Because the matched pair design yields a deterministic allocation plan, we do not present its results here. 

\begin{figure}[!h]
    \centering\spacingset{1.0}
    \includegraphics[width=1.0\textwidth]{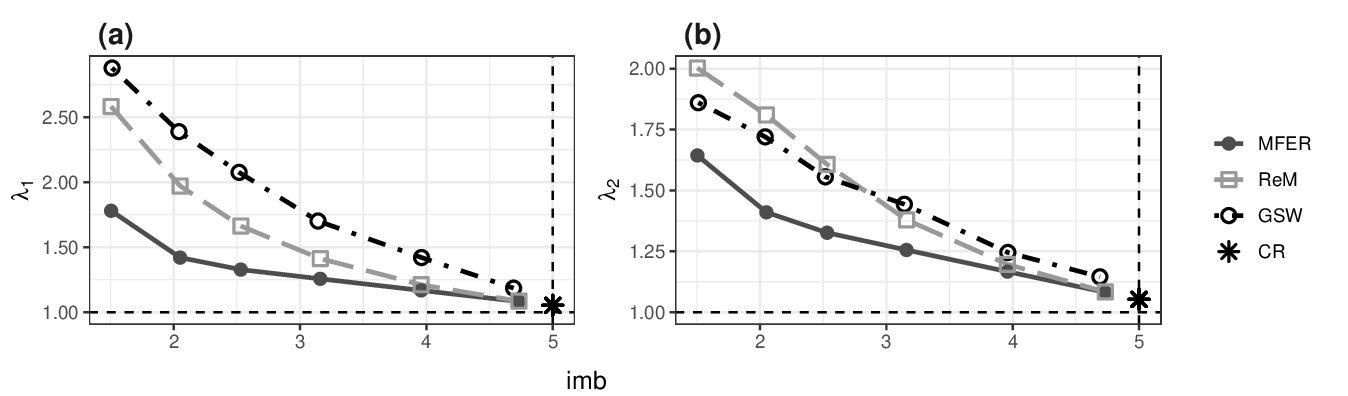}
    \caption{Average imbalance (imb) versus top two eigenvalues of $\E_{\bpi}\{(2W-1)(2W-1)^\T\}$ (from left to right) under $n=20$ against different methods. 
   The CR’s eigenvalue and average imbalance are shown by dotted-dashed lines for reference.
   }
    \label{fig:pareto_frontier}
\end{figure}

Figure \ref{fig:pareto_frontier} illustrates the trade-off between the imbalance measure and the top two eigenvalues of $\E_{\bpi}\{(2W-1)(2W-1)^\T\}$. The CR achieves the smallest top two eigenvalues, while the average imbalance is the largest. 
As the covariate imbalance is forced to decrease (moving left along the x-axis), ReM and GSW exhibit a severe spectral collapse, with their maximum eigenvalues increasing dramatically. In contrast, MFER consistently dominates the three methods. 
Benefits from smooth penalties rather than hard truncations, MFER achieves the same expected imbalance while maintaining a significantly lower top two eigenvalaues correspond to unfavourable MSE, effectively pushing the Pareto frontier closer to the CR baseline.
This echoes Theorem~\ref{thm:though} that under the same level of imbalance, MFER is the least harmful strategy because it shrinks not only the largest eigenvalue, but also the leading eigenvalues, making the unconstrained directions more isotropic.

{
\section{Real-Data Analysis: Blood Storage Duration and Prostate Cancer Recurrence}
\label{sec:real_data}
\citet{cata2011} conducted a randomized experiment to test whether transfusion of older red blood cells increases the risk of biochemical recurrence in prostate cancer patients. 
We reproduce this analysis using the corresponding dataset from the R package \texttt{medicaldata} \citep{medicaldata2021} and restrict the analysis to patients who experienced biochemical recurrence.
The response variable is the time to recurrence (\texttt{TimeToRecurrence}), and we binarize the treatment by storage duration. To be precise, we set $W=0$ for storage durations of $13$ days or less and $W=1$ otherwise. 
There are $14$ concomitant variables across four clinical domains: demographics, preoperative clinical metrics, postoperative pathological features, and adjuvant therapies. We treat all of them as $X$.

To synthetically generate potential outcomes, we first regress the observed outcome $\bY = (Y_1,\ldots, Y_n)^\T$ on the treatment indicator $W$ and the covariate $X$, obtaining the estimated coefficients ${\bbeta}$, a treatment effect of ${\tau}=12.5$, and $R_0^2\approx 0.43$.
We synthesize the potential outcomes as described in Section~\ref{sec:simulation} by replacing $\bbeta,\tau$ and $R_0^2$ with these estimators.
As a complement, we also evaluate a counterfactual based only on observed $Y_i$'s, \texttt{R4}, defined by $Y_i(W_i) = Y_i$ and $Y_i(1-W_i)=Y_i + (1 - 2W_i){\tau}$. 

\begin{table}[!h]
\centering\spacingset{1.0}
\caption{\label{tab:real_mse_main}Average imbalance (imb), max eigenvalue of $\E_{\bpi}\{(2W-1)(2W-1)^\T\}$ ($\lambda_{\max}$) and relative MSE under \texttt{R1}--\texttt{R4} among different methods. 
The CR and MP cannot control the imbalance and are reported for reference.
The optimal values across the remaining three covariate-adaptive designs are highlighted in bold.}
\centering
\setlength{\tabcolsep}{14pt}
\centering
\resizebox{\ifdim\width>\linewidth\linewidth\else\width\fi}{!}{
\begin{tabular}[t]{lrrrrrr}
\toprule
\multicolumn{3}{c}{ } & \multicolumn{4}{c}{MSE} \\
\cmidrule(l{3pt}r{3pt}){4-7}
\raisebox{3.25ex}[0pt][0pt]{Method} & \raisebox{3.25ex}[0pt][0pt]{imb} & \raisebox{3.25ex}[0pt][0pt]{$\lambda_{\max}$} & \multicolumn{1}{c}{\raisebox{1.5ex}{\small \texttt{R1}}} & \multicolumn{1}{c}{\raisebox{1.5ex}{\small \texttt{R2}}} & \multicolumn{1}{c}{\raisebox{1.5ex}{\small \texttt{R3}}} & \multicolumn{1}{c}{\raisebox{1.5ex}{\small \texttt{R4}}}\\[-2ex]
\midrule
MFER & 4.61 & \textbf{1.00} & \textbf{1.00} & \textbf{1.00} & \textbf{1.00} & \textbf{1.00}\\
ReM & 4.81 & 1.29 & 1.01 & 1.26 & 1.12 & 1.04\\
GSW & 4.62 & 1.23 & 1.04 & 1.26 & 1.14 & 1.03\\
MP & 6.42 & 1.04 & 1.19 & 1.25 & 1.38 & 1.04\\
CR & 13.99 & 0.55 & 1.17 & \multicolumn{1}{c}{-} & \multicolumn{1}{c}{-} & 1.24\\
\bottomrule
\end{tabular}}
\end{table}

In parallel to Table~\ref{tab:results}, the results of the blood storage study are reported in Table~\ref{tab:real_mse_main}.
As expected, MFER still performs best, as previously discussed. For inference purposes, we construct randomization‑based confidence intervals.
The exact randomization variance of the $\SATE$ is known in closed form as a function of \(\tau\):
\(V(\tau) = 4n^{-2}(\bY - W\tau)^\T \E_{\bpi}\{(2W-1)(2W-1)^\T\} (\bY - W\tau)\).
The \((1-\alpha)\) confidence bounds are then the roots of the quadratic inequality
\((\hat\tau - \tau)^2 \le z_{1-\alpha/2}^2\,V(\tau)\), where $z_{1-\alpha/2}$ denotes the $(1-\alpha/2)$th quantile of the standard normal distribution\citep[][Chapter~5]{imbens2015causal}. This yields coverage rates and interval lengths for every design.

\begin{table}[!h]
\centering\spacingset{1.0}
\caption{\label{tab:real_inf_main}Average confidence interval (CI) lengths and empirical coverage rates (in parentheses) under \texttt{R1}--\texttt{R4} across different methods. The shortest CI lengths among the covariate-adaptive designs are highlighted in bold.}
\centering
\resizebox{\ifdim\width>\linewidth\linewidth\else\width\fi}{!}{
\begin{tabular}[t]{lrrrr}
\toprule
\multicolumn{1}{c}{ } & \multicolumn{4}{c}{CI length (coverage rate)} \\
\cmidrule(l{3pt}r{3pt}){2-5}
\raisebox{3.25ex}[0pt][0pt]{Method} & \multicolumn{1}{c}{\raisebox{1.5ex}{\small \texttt{R1}}} & \multicolumn{1}{c}{\raisebox{1.5ex}{\small \texttt{R2}}} & \multicolumn{1}{c}{\raisebox{1.5ex}{\small \texttt{R3}}} & \multicolumn{1}{c}{\raisebox{1.5ex}{\small \texttt{R4}}}\\[-2ex]
\midrule
MFER & \textbf{20.99} (0.951) & \textbf{25.20} (0.997) & \textbf{22.69} (0.958) & \textbf{20.44} (0.951)\\
ReM & 21.11 (0.951) & 28.02 (0.995) & 23.87 (0.959) & 20.87 (0.951)\\
GSW & 21.44 (0.951) & 28.11 (0.996) & 24.11 (0.959) & 20.82 (0.953)\\
MP & 23.35 (0.953) & 29.30 (0.951) & 27.39 (0.953) & 21.36 (0.953)\\
CR & 22.52 (0.951) & \multicolumn{1}{c}{-} & \multicolumn{1}{c}{-} & 22.47 (0.948)\\
\bottomrule
\end{tabular}}
\end{table}

Table~\ref{tab:real_inf_main} reports the average length of the 95\% confidence interval together with the coverage rate. 
As highlighted in the table, the proposed MFER attain the shortest lengths among covariate‑adaptive designs.

}
}

\bibliographystyle{chicago}
\bibliography{report}

\end{document}